\documentclass{appolb}
\usepackage{epsfig}

\begin{document}

\title{Mirror dark matter discovered?}
\author{ Z.K. Silagadze
\address{Budker Institute of Nuclear Physics \\ and \\ Novosibirsk State
University \\ 630 090, Novosibirsk, Russia}
}
\maketitle
\begin{abstract}
Recent astrophysical data indicates that dark matter shows a controversial 
behaviour in galaxy cluster collisions. In case of the notorious Bullet
cluster, dark matter component of the cluster behaves like a collisionless
system. However, its behaviour in the Abell 520 cluster indicates 
a significant self-interaction cross-section. It is hard for the WIMP based 
dark matter models to reconcile such a diverse behaviour. Mirror dark matter 
models, on the contrary, are more flexible and for them diverse behaviour 
of the dark matter is a natural expectation. 
\end{abstract}

\section{Introduction}
The evidence of dark matter is compelling at all observed astrophysical 
scales \cite{1-1,1-2}. At that the dark matter reveals itself only through 
its gravitational interactions. There are no observational facts that
indicate existence of any significant interactions between the ordinary
and dark matter particles. However, some recent astrophysical observations
show that dark matter particles may have significant self-interactions. This
is surprising because other observations support the WIMP paradigm that the
dark matter is essentially collisionless. The mirror dark matter \cite{1-3,
1-4,1-5}, first introduced in \cite{1-KOP}, has a potential to reconcile 
these seemingly contradictory observational facts. For other attempts to do 
this, on the base of BEC dark matter, see \cite{1-6}. Below, after briefly 
commenting the mirror matter idea, we consider the above mentioned 
observational facts and argue why, in our opinion, they indicate the existence 
of the mirror matter as the main dark matter component.

\section{Mirror matter}
One of characteristic features of our present day  increasingly shallow 
postmodern culture is the existence of a ``literature horizon'' beyond which 
many interesting results of the past have fallen. ``References familiar to 
one generation are often less so or unknown to the successive one, an effect
which increases with the passage of time'' \cite{2-1}.

It is still widely known that in their famous article \cite{2-2} Lee and 
Yang revealed  ``The fact that parity conservation in the weak interactions
was believed for so long without experimental support'' \cite{2-3} and 
hypothesized the possibility of parity non-conservation in the weak
interactions. The hypothesis turned out to be true, as the subsequent
experiments had shown, and nowadays it is a firmly established fact that our
universe is left-handed as far as the weak interactions are concerned.

What is not so widely known is the fact that at the end of the very same
paper Lee and Yang indicated how left-right symmetry of the world could be
rescued by duplicating the non-symmetric part of our left-handed universe in 
the mirror.

It is a futile business to establish priority in the genesis of physical
ideas. Seldom a good physical idea is already mature at the time of its
first appearance and, as a rule, many people participate in its subsequent 
developments.

In a very postmodern fashion, we can deprive Lee and Yang priority of the
mirror matter idea and attribute it, for example, to Lewis Carroll. To
paraphrase his {\it Alice Through the Looking Glass}, ``the mirror particles 
are something like our particles, only the chiralities go the wrong way 
\ldots '' \cite{2-4}.

Jorge Luis Borges, famous Argentine poet, essayist, and short-story writer,
with his classic tales of fantasy and dreamworlds, is another candidate to 
whom the idea of duplication of the world through mirrors could be attributed.
Berezinsky and Vilenkin cite \cite{2-5} somewhat incomplete quote from 
Borges: ``The visible universe is an illusion. Mirrors \ldots are hateful 
because they multiply it''. We found that it appears in the short story of 
Borges {\it Tl\"on, Uqbar, Orbis Tertius}, written in 1940, and in the 
complete form has somewhat different accents,  going like this: ``For one of 
those gnostics, the visible universe was an illusion or (more precisely) 
a sophism. Mirrors and fatherhood are abominable because they multiply and 
disseminate that universe'' \cite{2-6}. 

\begin{figure}[htbp]
\begin{center}
\epsfig{figure=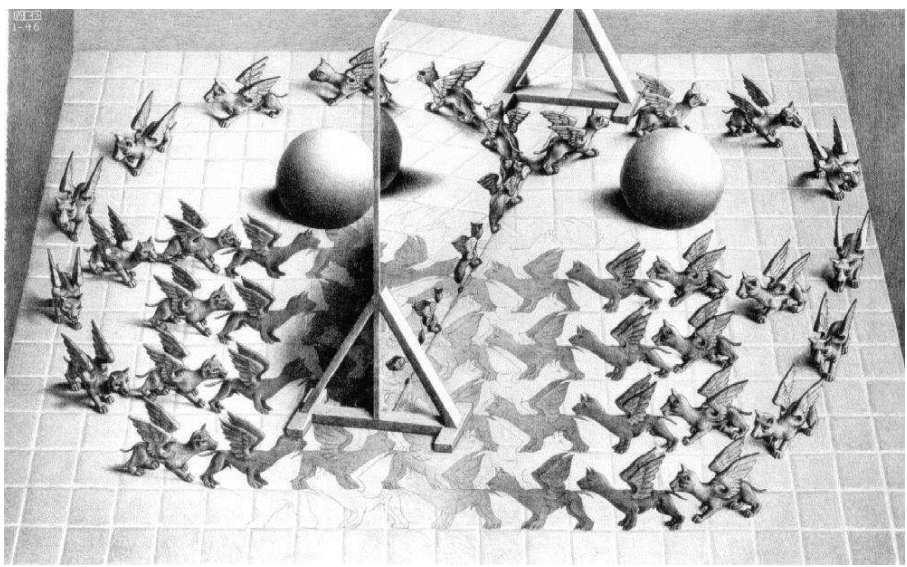,width=12cm}
\end{center}
\caption {Mirror matter idea illustrated by M.~C.~Escher's lithograph `Magic
Mirror'.}
\label{Fig1}
\end{figure}

However, the most complete and clear explanation of the mirror matter idea is
provided by 1946 lithograph print ``Magic Mirror'' of famous Dutch artist 
M.~C.~Escher (see Fig.\ref{Fig1}). Our world is presented at this lithograph
as a left-handed procession of small griffins (winged lions). Griffins are
symbols of ordinary particles which participate in the left-handed (V-A)
weak interactions. Interactions between ordinary particles are governed by
the Standard Model gauge group $G=SU(3)\times SU(2)\times SU(1)$ (or by its 
Grand Unification or supersymmetric extension). The corresponding gauge bosons
are indicated at the lithograph, perhaps, by a small solid sphere around
which the griffins' procession goes on. This part of the universe (lithograph)
is clearly not mirror-symmetric (parity is violated in our world). However,
when reflected in the mirror it gives a birth of the mirror partners of
every ordinary particle including the gauge ones. Mirror griffins
participate in a right-handed procession around the mirror copy of the
the Standard Model gauge group $G^\prime=SU^\prime(3) \times SU^\prime(2)
\times SU^\prime(1)$. Mirror weak interactions are right-handed (V+A) and in
overall the mirror symmetry of the universe (lithograph) is restored.

As we see, the idea of Mirror World was already presented in fiction
literature and art when Lee and Yang gave the beginning of the physical
incarnation of the idea which immediately had fallen beyond the literature
horizon. After a decade, Kobzarev, Okun and Pomeranchuk, influenced by the
Landau's idea of combined parity and subsequent experimental discovery of 
{\bf CP}-violation, digged out the mirror matter concept from the literature 
horizon and gave it the first serious phenomenological consideration 
\cite{1-KOP}. This paper marked the real beginning of the mirror 
matter story. It was argued that almost all elementary particles should
be duplicated (with possible exclusion of some neutral particles) and that
the mirror particles can not have common strong and electromagnetic 
interactions with ordinary particles. The concepts of ``Mirror Matter'' and
corresponding invisible ``Mirror World''(after Lewis Carroll's  {\it Alice 
Through the Looking Glass}), as complex as our own world, were introduced in 
this work for the first time and observational effects of the mirror matter 
were investigated. Some further investigations of possible astrophysical 
effects of the hidden sector particles followed \cite{2-7,2-8,2-9}, but the 
idea was still not far from the literature horizon until it was rediscovered 
in the modern context of renormalizable gauge theories by Foot, Lew and Volkas 
\cite{2-10} and used in the context of neutrino oscillations \cite{2-11,2-12}. 

Okun's recent review \cite{2-13} cites more than 250 references
related to the mirror matter idea and we hope that it remains outside the 
literature horizon. However the idea is still little known to the majority 
of physicists, oriented at the mainstream.

\section{Empirical evidence of dark matter from the bullet cluster}
Galaxy clusters are the largest gravitationally constrained structures sitting
atop of the hierarchical structure formation in the universe \cite{3-1,3-2}.
Because of their size, their matter content resembles the matter content of
the Universe with the dark matter as the major mass component (about 85\%). 
Two other main ingredients of a typical cluster are stars which are grouped
to form member galaxies (about 5\% of the total cluster mass) and
very hot (virialized) intergalactic gas (about 10\%).

This three components behave differently when two galaxy clusters collide.
Note that merger events are typical for hierarchical structure formation and
drive the formation of larger systems from the smaller ones. However, at the
scale of galaxy clusters mergers are rather rare events. Nevertheless several
interesting merger events were observed recently which constitute fascinating
laboratories to study dark matter behavior under these titanic collisions.
Most dark matter theories predict that self-interaction of the dark matter 
particles (WIMPs) are very feeble and, therefore, dark matter is 
collisionless.
 
Then it is expected that the dark matter components of the colliding clusters 
do not perturb much each other during the collision allowing them to pass 
right through. The stars in the member galaxies also form collisionless system 
because stars are sparse and star collisions are rare. Therefore, during the 
collision it is expected dark matter to follow the galaxies and these two 
components of the cluster should stay together even after the collision, while 
the third component (hot intergalactic gas) having strong electromagnetic 
self-interaction should lag behind near the collision center and show signs of 
hydrodynamic disturbances like shock waves formed as a result of supersonic 
collisions of gas components of colliding clusters.

\begin{figure}[htbp]
\begin{center}
\epsfig{figure=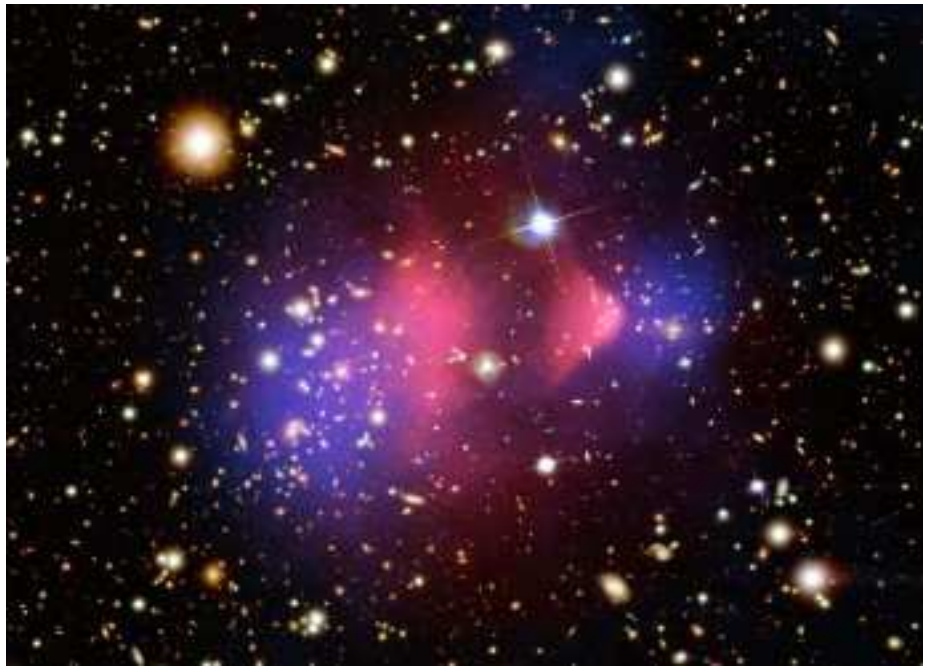}
\end{center}
\caption {Dark matter in the Bullet cluster.}
\label{Fig2}
\end{figure}

This is precisely what is observed in the Bullet Cluster, more formally known 
as 1E0657-56. Fig.\ref{Fig2} is a composite image of this cluster. The Hubble
Space Telescope and Magellan optical image shows cluster's member galaxies in 
orange and white. The distribution of hot intergalactic gas is traced by 
Chandra X-ray observations and is superimposed on the image in pink. The dark 
matter component of the cluster can not be seen neither in optical nor in
X-rays, but it can be revealed by gravitational lensing which leads to the 
distortion of images of background galaxies. The distribution of the dark 
matter (unseen concentration of mass) found by gravitational lensing studies 
is indicated in blue in  Fig.\ref{Fig2}. 

We see that the galaxies and the corresponding dark matter clumps stay 
together and the pink clumps lag behind. Besides, a classic bow-shaped shock 
wave is clearly seen on the right  in the pink image of the smaller cluster 
hot gas component giving it a  bullet-like appearance.

We can conclude, therefore, that the Bullet cluster constitutes the first
direct empirical evidence in favor of dark matter \cite{3-3}.  At first 
sight, all these are good news for WIMP based dark matter models. Especially
if we take into account that the similar behavior was also observed in other
merging galaxy clusters CL 0152-1357 \cite{3-4} and MS 1054-0321 \cite{3-5}.
In latter case the dark matter clumps seem to be offset not only from the
corresponding X-ray peaks but also from the galaxy counterparts, as if they 
were moving ahead of the cluster galaxies. Therefore, in this particular case,
the dark matter behaves as effectively more collisionless than the cluster 
galaxies. Another interesting peculiarity of MS 1054-0321 is that there is no
associated X-ray peak in the eastern region where the weak-lensing and 
luminosity map reveal significant mass concentrations \cite{3-5}. This maybe
indicates that most of the intracluster gas of the eastern clump has been 
stripped by RAM pressure while passing through the denser regions of the 
colliding cluster \cite{3-5}. 
		
\section{Dark matter in the Abell 520 cluster}
\begin{figure}[htbp]
\begin{center}
\epsfig{figure=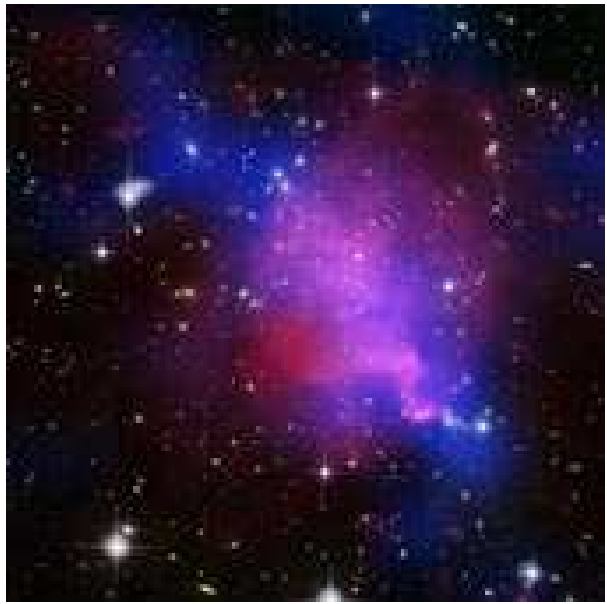,width=8cm}
\end{center}
\caption {Dark matter in the Abell 520 cluster.}
\label{Fig3}
\end{figure}

Matters are not so simple, however. Observations of another merging galaxy
cluster Abell 520 (see Fig.\ref{Fig3}) seem to be puzzling for WIMP dark 
matter theories \cite{4-1}.

As expected, X-ray emission is offset from the 
galaxy distribution. However, in contrast to the Bullet cluster, the lensing 
signal and the X-ray emission coincide and lag behind the galaxies indicating
that the dark matter is collisional like the ordinary gases. The inferred 
estimate on the dark matter self-interaction cross section is well above the
upper limit derived for the Bullet cluster and exceeds by many orders the 
cross section magnitude expected for WIMPs \cite{4-1}. 

Certainly, mirror dark matter ``is richer than the dark matter of SUSY'' 
\cite{2-13} and has a better chance to be reconciled with these contradictory
astrophysical observations.

It is important to realize that at macroscopic level mirror and ordinary 
matters are expected to behave differently \cite{4-2,4-3} because the 
nucleosynthesis bounds demand the mirror sector to have a smaller 
temperature than the ordinary one and hence a different cosmological 
evolution.

In particular, mirror stars should be helium dominated and evolve faster than 
the ordinary ones \cite{4-4}. Consequently, mirror supernova rate is expected
to be larger and star formation more efficient in the mirror sector due to  
shock waves related to the supernova explosions. As a result, we expect mirror
galaxies to contain less gas compared to the ordinary galaxies \cite{4-4}.
Nevertheless some fraction of mirror gas will be inevitably present.

In case of the mirror dark matter, therefore, we expect diverse behavior
during galaxy cluster collisions, as diverse as for the ordinary matter.
At that a typical cluster of mirror galaxies is expected to be less gas 
dominated than its ordinary counterpart leading, therefore, to the Bullet
cluster like behavior. However, other atypical possibilities like Abell 520 
is also not excluded.  

\section{Further evidence that dark matter behaves like ordinary matter}
\begin{figure}[htbp]
\begin{center}
\epsfig{figure=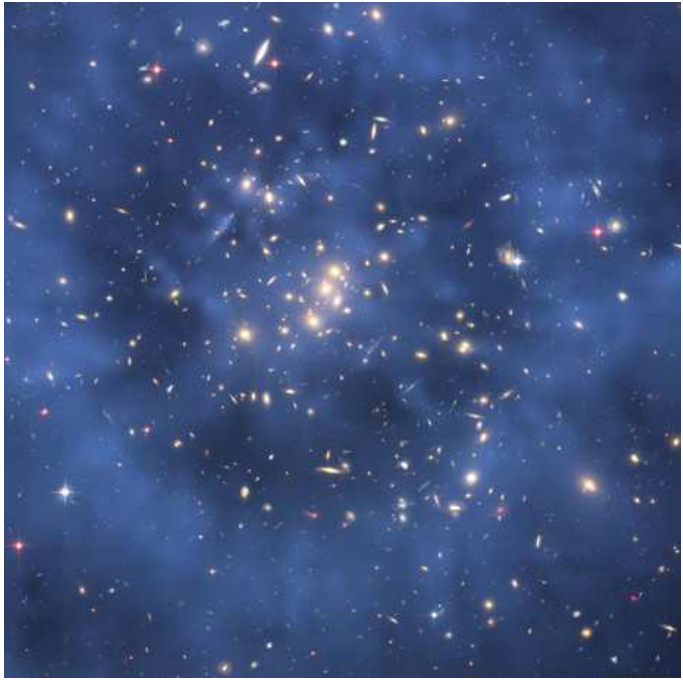,width=9cm}
\end{center}
\caption {Dark matter ring in Cl 0024+17 galaxy cluster.}
\label{Fig4}
\end{figure}

Could we find other indications that dark matter behaves like ordinary matter
forming a diversity of structures? One possible candidate is a remarkable
discovery of a ringlike dark matter structure in the galaxy cluster 
Cl 0024+17 \cite{5-1}. This galaxy cluster is like the Bullet cluster 
but viewed along the collision axis at a much later epoch \cite{5-1}. 
Fig.\ref{Fig4} is a composite image of Cl 0024+17 with reconstructed
distribution of the dark matter indicated in blue. A huge dark matter ring 
of 2.6 million light-years across is clearly seen.

For ordinary matter, ringlike structures are common output of violent 
collision events both at the scale of galaxies and at the scale of galaxy
clusters. For example, Very Large Array radio telescope discovered recently
a giant ring-shaped radio-emitting structures of about 6 million light-years 
across around the galaxy cluster Abell 3376 \cite{5-2} probably resulting 
from the megaparsec scale cosmic shock waves associated to the violent 
collisions of smaller groups of galaxies within  the cluster. 

A mechanism how ringlike structures can be formed during collisions was 
suggested by Lynds and Toomre \cite{5-3} many ears ego. When a compact 
intruder approaches with a small impact parameter in a nearly head-on 
collision to a victim galaxy, the stars of this galaxy, assumed to be in 
circular orbits before the collision, experience inward fall due to extra 
gravity. As the intruder moves away, extra gravity disappears and perturbed
stars rebound outward eventually. At that stars at smaller radii rebound 
faster and, as a result, stars still moving inward meet rebounders moving 
outward. The final outcome all of this is a ring-shaped compression wave 
propagating outward \cite{5-4,5-5}.

If the dark matter ring in Cl 0024+17 is created only because of 
gravitational disturbances during the collision, in the manner of Lynds and 
Toomre, it is expected the spatial distribution of cluster galaxies to 
possess the similar ringlike feature. However a detailed dedicated study 
found no such substructure in the projected two-dimensional galaxy 
distribution \cite{5-6}. This indicates that probably dark-matter 
self-interactions and associated shock waves played an important role in the 
formation of the observed dark matter ring.

On the other hand, the fact that the ring has not been erased in about 
1-2~Gyr after the core impact indicates very small collisional  cross-sections 
of dark matter particles, much smaller than expected for ordinary plasma
\cite{5-1}.

The mirror dark matter hypothesis readily explains this seeming contradiction 
in the behavior of the Cl 0024+17 cluster dark matter. The shock wave 
propagated in the mirror gas of the cluster leads to the intense burst of 
mirror star formation transforming the dark matter ring into the effectively
collisionless system of mirror galaxies.
   
This example also indicates how the ordinary and mirror matters could to be 
separated even at galactic scales: mirror galaxies in the dark matter ring
of Cl 0024+17 should be composed predominantly by mirror matter. The existence
of purely dark matter galaxies with negligible admixture of ordinary matter, 
as well as ordinary galaxies without dark matter, can therefore be envisaged.
Interestingly, both dark galaxy \cite{5-7,5-8} and ordinary galaxy without
dark matter \cite{5-9} were presumably discovered.
 
\section{Hoag's object}
\begin{figure}[htbp]
\begin{center}
\epsfig{figure=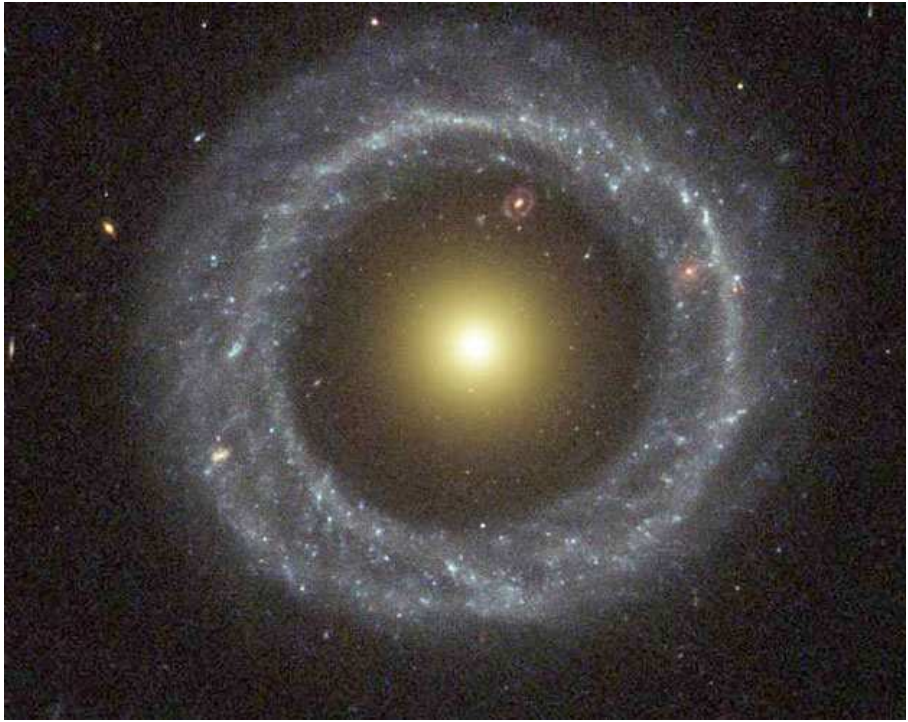,width=12cm}
\end{center}
\caption {Hoag's object.}
\label{Fig5}
\end{figure}

Fig.\ref{Fig5} shows a very interesting ring galaxy, the so called Hoag's
object \cite{6-1}. Yellow spherical nucleus of old stars is surrounded by 
a nearly perfect ring of hot blue stars. How this ring was formed is a
mystery. Usually ring galaxies are formed by the collision of a small galaxy 
with a larger disk-shaped galaxy through the Lynds and Toomre density wave
mechanism. However, in the case of the Hoag's object there is no sign of any
intruder galaxy nearby.

Curiously, an object is seen inside the Hoag's object's image which looks like
a another ring galaxy much like to the Hoag's object itself. Having in mind
the estimated minuscule fraction, $10^{-3}$, of Hoag-type galaxies 
\cite{6-2}, this seems to be an incredible coincidence. Maybe this odd
cosmic irony offers a clue about the nature and formation history of the
Hoag's object.

Suppose the intruder galaxy which led to the ring formation was a mirror
galaxy. Then we do not expect it to be visible but it should be lurking
somewhere there and can reveal itself by its gravitational influence,
in particular by gravitational lensing of background galaxies suitably
aligned behind it \cite {6-3,6-4}.

What if the smaller ring-galaxy-like structure inside the Hoag's object
is the result of the gravitational lens of some distant background galaxy
due to the gravitational field of the mirror galaxy?

Note that Hoag himself considered a hypothesis that the ring of the Hoag's 
object is a gravitational lensing event caused by the gravity of Hoag's
object's core \cite{6-1}. The hypothesis was discarded because it required
inordinately high mass for the core of the Hoag's object, $1.4\times10^{13} 
~M_{\odot}$ \cite{6-1,6-5}, while the inferred mass of the Hoag's object 
is only $7^{+5}_{-3}\times10^{11}~M_{\odot}$ \cite{6-2}.

Nowadays we firmly know that the Hoag's object's ring is real, of course, 
but one cannot be so sure about the much smaller duplicate ring inside. 
The apparent size of the secondary ring is a factor of twenty smaller than 
the size of the main ring. The radius of the Einstein ring is proportional 
to the square root of the lensing mass \cite{6-4}. Therefore, the required 
lensing mass which can produce the Einstein ring of the angular size of the 
secondary ring is $1.4\times10^{13}~M_{\odot}/20^2\approx 3.5\times 
10^{10}~M_{\odot}$, maybe just a correct mass for the putative projectile 
for the Hoag's object formation.
 
\section{Conclusions}
There is a growing astrophysical evidence of diverse behaviour of dark 
matter in galaxy cluster collisions. These observations indicate that
the mirror dark matter models deserve careful examination and observational
verification. If the mirror dark matter were as popular as the SUSY dark 
matter, we would say that it is already discovered. However, it would be
more fair to conclude that we need more observational evidence to firmly 
prove this fascinating conjecture.

\section*{Acknowledgments}
The author thanks L.~B.~Okun for comments.
The work is supported in part by grants Sci.School-905.2006.2 and 
RFBR 06-02-16192-a.

\end{document}